\documentclass[11pt,a4paper]{article}
\usepackage{jheppub,graphicx,amsmath,amssymb}

\newcommand{\nn}{\nonumber}
\newcommand{\Mpl}{\overline{M}_{\rm Pl}}
\newcommand{\gro}{\tilde G}
\newcommand{\go}{\tilde g}
\newcommand{\Etmiss}{\not\hspace{-1.2mm}E_{T}}

\title{
Light gravitino production in association with gluinos at the LHC
}

\author[a,b]{P.~de Aquino,} 
\emailAdd{priscila@itf.fys.kuleuven.be} 
\author[b]{F. Maltoni,}
\emailAdd{fabio.maltoni@uclouvain.be} 
\author[c]{K.~Mawatari}
\emailAdd{kentarou.mawatari@vub.ac.be}
\author[c]{and B.~Oexl}
\emailAdd{bettina.oexl@vub.ac.be}

\affiliation[a]{Instituut voor Theoretische Fysica, 
 Katholieke Universiteit Leuven,\\ 
 Celestijnenlaan 200D, B-3001 Leuven, Belgium} 
\affiliation[b]{Centre for Cosmology, 
 Particle Physics and Phenomenology (CP3),\\
 Universit\'e Catholique de Louvain, B-1348 Louvain-la-Neuve, Belgium}
\affiliation[c]{Theoretische Natuurkunde and IIHE/ELEM, 
 Vrije Universiteit Brussel,\\
 and International Solvay Institutes, Pleinlaan 2, 
 B-1050 Brussels, Belgium}

\abstract{
We study the jets plus missing energy signature at the LHC in a scenario
where the gravitino is very light and the gluino is the next-to-lightest
supersymmetric particle and promptly decays into a gluon and a
gravitino. We consider both associated gravitino production with a
gluino and gluino pair production. By merging matrix elements with
parton showers, we generate inclusive signal and background samples and
show how information on the gluino and gravitino masses can be obtained
by simple final state observables. 
}

\begin{document}
\maketitle

\section{Introduction}\label{sec:intro}

Identification and interpretation of new physics signals are formidable
challenges at the LHC. A promising and rather general signature for
probing new physics at hadron colliders is jets plus missing transverse
energy ($\Etmiss$)~\cite{Morrissey:2009tf}. Particularly, the signature
has been commonly studied in the context of the minimal supersymmetric
extension of the Standard Model (SM) and also simplified models with
the lightest neutralino as the lightest supersymmetric particle
(LSP). In these models, strongly-interacting superpartners (gluinos
and/or squarks) are copiously produced and decay into gluons/quarks and
stable neutralinos, leading to a signal containing multiple jets and
missing energy. So far no excess over the SM background expectation has
been observed at the 
Tevatron~\cite{Abachi:1995ng,Affolder:2001tc,Abazov:2007aa,Aaltonen:2008rv}
and the LHC~\cite{Chatrchyan:2011zy,Aad:2011ib}, which is interpreted as
exclusion limits in the traditional $m_0-m_{1/2}$ plane or in the
gluino$-$squark mass plane. On the other hand, another interesting LSP
candidate is the gravitino, and such scenarios have not been fully
explored in the jets+$\Etmiss$ signature.%
\footnote{The gravitino LSP scenarios have often been searched in
diphoton events with missing energy for the lightest neutralino as the
next-to-lightest supersymmetric particle
(NLSP)~\cite{Aaltonen:2009tp,Abazov:2010us,Aad:2011zj,Chatrchyan:2011wc}.} 
This is the primary target in this article. 

Gravitinos are the spin-3/2 superpartners of gravitons and become
massive via the super-Higgs mechanism by absorbing massless spin-1/2
goldstinos~\cite{Deser:1977uq,Cremmer:1978iv,Cremmer:1982wb}. While the
interactions of the helicity 3/2 components of the gravitino are
suppressed by the Planck scale, those of the helicity 1/2 components are
suppressed by the SUSY breaking scale if the gravitino mass is much
smaller than the energy scale of the interactions. Therefore, if the
SUSY breaking scale is low, the gravitino interactions, i.e. the
goldstino interactions, can be important at colliders. We also note
that, as a consequence of the super-Higgs mechanism, the gravitino mass 
is related to the scale of SUSY breaking as well as the Planck scale,
\begin{align}
 m_{3/2}\sim (M_{\rm SUSY})^2/M_{\rm Pl}.
\end{align}
Therefore, low-scale SUSY breaking scenarios, e.g. gauge mediated SUSY
breaking (GMSB) \cite{Giudice:1998bp}, provide a gravitino LSP. 

Several studies of the jets+$\not\hspace{-1.8mm}E_{T}$ signature in
gravitino production 
at hadron colliders have been performed especially for very light
gravitinos with   
$m_{3/2}\sim{\cal O}(10^{-14}-10^{-12}$~GeV)~\cite{Dicus:1989gg,Drees:1990vj,Dicus:1996ua,Kim:1997iwa,Brignole:1998me,Klasen:2006kb}.
In such a very light gravitino case, production in
association with a gluino (or squark) can be dominant or comparable to
usual gluino pair production.%
\footnote{Reference~\cite{Brignole:1998me} assumes that all SUSY
particles except for the gravitino are too heavy to be produced
on-shell.} 
The subsequent gluino decays into a gluon and a
gravitino will give rise to monojet and dijet signals with missing
energy. The associated gravitino production is
significant only for very light gravitinos since the production rate is
inversely proportional to the square of the Planck scale times the 
gravitino mass, $\sigma\propto 1/(M_{\rm Pl\,}m_{3/2})^2$. The case for
the gravitino mass of $m_{3/2}\sim{\cal O}(10^{-9}$~GeV) in GMSB
has also been studied in gluino NLSP
scenarios~\cite{Baer:1998pg,Kats:2011qh}, where the
dijet+$\not\hspace{-1.8mm}E_{T}$ signal can be significant. No realistic
study including parton shower and hadronization effects, however, has
been conducted, mainly due to the limited availability of simulation
tools for processes involving gravitinos. 

To be able to identify new physics in such a multi-jet signature at the
LHC, a reliable and precise simulation of the signal as well as of the QCD
background is crucial. This can be provided by merging matrix elements
(ME) with parton showers (PS). In the last decade several techniques to
consistently merge multi-parton final states as obtained by a 
ME computation with PS (ME+PS) 
have been developed~\cite{Alwall:2007fs}. They are now implemented in 
various event generators, and tested against experimental
data (see~\cite{Buckley:2011ms} for a review). Moreover, the importance
of the ME+PS merging for new physics has been pointed out in different 
contexts~\cite{Plehn:2005cq,Alwall:2008ve,Alwall:2008va,Alwall:2008qv,Plehn:2008ae,deAquino:2011ix}.  

In this article we consider a scenario where the gravitino is the LSP
and the gluino is the NLSP and promptly decays into a gluon and a
gravitino ($\go\to g+\gro$). We study the jets plus missing transverse
energy signature
\begin{align}
 pp\to\text{jets}\ +\Etmiss,
\label{signal}
\end{align}
where the missing energy is due to two gravitinos. For simplicity, all
other superparticles are assumed to be too heavy to be produced
on-shell. We consider a very light gravitino case, where two main
production mechanisms contribute to the above signal: gluino-gravitino
associated production ($pp\to\go\gro$) and gluino pair production
($pp\to\go\go$), to be described in detail in
section~\ref{sec:subprocesses}. Thanks to the availability of new
simulation tools it is now possible to apply the ME+PS merging procedure
to avoid double counting for such a signal which contains two different
types of subprocesses.%
\footnote{A similar issue might arise in neutralino LSP scenarios when
$\tilde q\to q+\tilde\chi_1^0$. $\tilde q\tilde\chi_1^0$ associated 
production leads to monojet+$\Etmiss$, while $\tilde q\tilde q$
production gives dijet+$\Etmiss$. However, the $\tilde q\tilde\chi_1^0$
production rate is much smaller than the $\tilde q\tilde q$ production
due to the weak gauge coupling~\cite{Allanach:2010pp}.}
We generate the merged inclusive signal samples as well as the SM
background sample, and analyse the distributions of the jets and missing
transverse energy to extract information on the gluino and gravitino
masses.  

The article is organized as follows.
In section~\ref{sec:subprocesses}, the two production subprocesses
contributing to the jets plus missing energy at the LHC are presented,
i.e. gluino-gravitino associated production and gluino pair production.  
In section~\ref{sec:merging}, we briefly describe the matrix element and
parton shower merging technique employed in this work and the validation
of our generation.  
In section~\ref{sec:analysis} we examine basic selection cuts to curb
the SM background, and discuss how distributions of the jets and missing
transverse energy can be used to determine the SUSY particle masses. 
Section~\ref{sec:summary} is devoted to our summary. 
In appendix~\ref{sec:lagrangian}, we give the effective gravitino
interaction Lagrangian relevant to our study.

\section{Light gravitino production at the LHC}
\label{sec:subprocesses}

We investigate LSP gravitino production processes in $R$-parity
conserving scenarios that lead to jets+$\not\hspace{-1.8mm}E_{T}$ at the
LHC. We consider gluinos to be the NLSP and to promptly decay into a
gluon and a gravitino.  We assume the masses of all other SUSY particles
large enough to prevent them from being produced on-shell. The missing
energy will be carried by two gravitinos due to the $R$-parity
conservation, and two processes,  
{\it gluino-gravitino associated production} and 
{\it gluino pair production}, whose importance varies with the gravitino
and gluino masses, can contribute to the final state
\begin{align}
 pp\to\text{partons}+\gro\gro.
\label{signal_parton}
\end{align}
Before considering the two processes in detail, we remark that
gravitino pair production ($pp\to\gro\gro$), where a graviton and the
scalar superpartners of the goldstinos (the so-called sgoldstinos)
$s$-channel exchange diagrams and the $t,u$-channel gluino exchange 
diagrams are involved, as well as sgoldstino pair production, might give
rise to the jets+$\Etmiss$ signal when extra QCD radiation is significant.%
\footnote{The gravitino pair and the sgoldstino pair production in
photon-photon collisions ($\gamma\gamma\to$ ``nothing'') were studied
in~\cite{Bhattacharya:1988ey}, while the inverse processes
($\gro\gro\to\gamma\gamma/f\bar f$) were investigated
in~\cite{Gherghetta:1996fm}. We also note that
ref.~\cite{Brignole:1998me} studied $\gro\gro g/q$ final states at
hadron colliders by means of the effective Lagrangian approach and the
collinear approximation, where all the SUSY particles except gravitinos
are assumed to be heavy.} 
However, those contributions are suppressed by the SUSY breaking scale
squared and the signal events can be expected only in the low $p_{T}$
region. Therefore, we expect them to be negligible in our signal region
and we do not include them in this work.

\subsection{Gluino-gravitino associated production}
\label{sec:go_gro_assoc}

Gravitino production associated with a gluino and the subsequent gluino
decay,
\begin{align}
 pp\to\go\gro\to g\gro\gro,
\label{process_gogro}
\end{align}
arises from the $gg$ and $q\bar q$ initial states, and leads to a
mono-jet plus missing energy signal at the leading order (LO). The
partonic cross section can be computed by using the effective gravitino
interaction Lagrangian, given in appendix~\ref{sec:lagrangian}, and the
analytic expression can be found, e.g., in~\cite{Klasen:2006kb}.%
\footnote{The analytic helicity amplitudes for $q\bar q\to\go\gro$ is
also available in~\cite{Mawatari:2011cu} after some substitutions for
the masses and couplings in the $e^+e^-\to\tilde\chi^0_1\gro$ process.} 
The cross section for the process is inversely proportional to the
square of the Planck scale times the gravitino mass 
\begin{align}
 \sigma(pp\to\go\gro)\propto 1/(M_{\rm Pl}\,m_{3/2})^2,
\label{Sigmagrogo}
\end{align}
and therefore it becomes significant at colliders only when the
gravitino is very light, $m_{3/2}\sim {\cal O}(10^{-12}$ GeV) or
less. As expected, gravitino production associated with other SUSY
particles also follows the scaling of eq.~\eqref{Sigmagrogo}. The
current experimental bound on the gravitino mass is given by the
mono-photon plus missing-energy signal in neutralino-gravitino
associated production at the LEP as a function of the neutralino and
selectron masses, e.g.  
\begin{align}
 m_{3/2}\gtrsim 10^{-14}\ {\rm GeV},
\label{massbound}
\end{align}
for $m_{\tilde{\chi}^0_1}=140$ GeV and 
$m_{\tilde e}=150$~GeV~\cite{Abdallah:2003np}. At the Tevatron a similar
bound on the gravitino mass is set from the 
$\gamma+\!\Etmiss$~\cite{Acosta:2002eq} and 
jet+$\Etmiss$~\cite{Affolder:2000ef} channels, where it is assumed that
all the other SUSY particles are too heavy to be produced
on-shell~\cite{Brignole:1998me}.   

\begin{figure}
\center
 \includegraphics[width=.5\textwidth,clip]{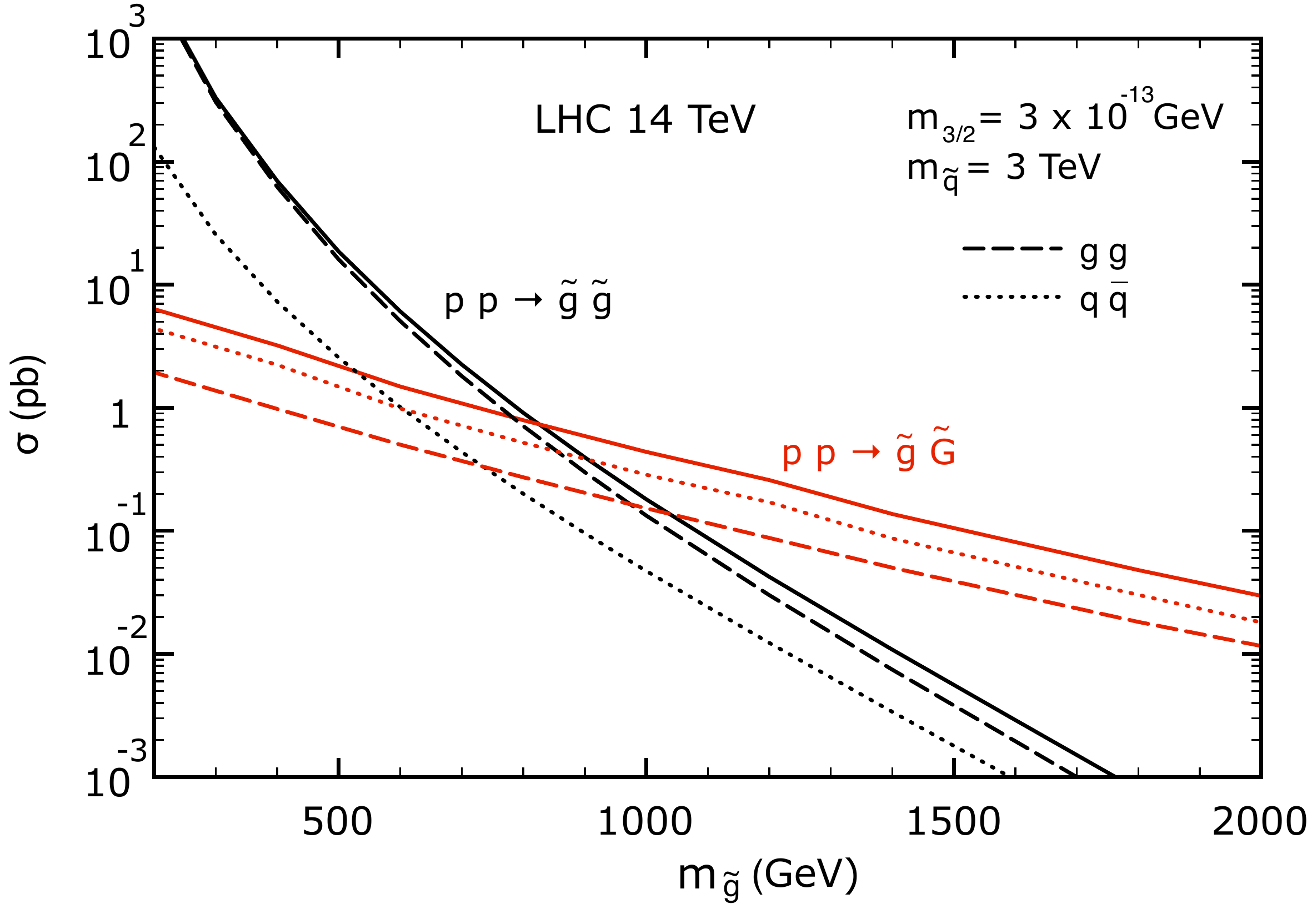} 
 \caption{Total cross sections of the gluino-gravitino associated
 production for the gravitino mass $m_{3/2}=3\times 10^{-13}$ GeV,
 $pp\to\go\gro$ (red), and the gluino pair production, $pp\to\go\go$
 (black), at the LHC with $\sqrt{s}=14$~TeV as a function of the gluino
 mass. The dashed and dotted lines represent the contributions of the
 $gg$ and $q\bar q$ initial states, respectively. The squark masses are
 fixed at 3 TeV.} 
\label{fig:xs_go_mass}
\end{figure}

Figure~\ref{fig:xs_go_mass} presents the total cross section of the
gluino-gravitino associated production \eqref{process_gogro} for
$m_{3/2}=3\times 10^{-13}$ GeV at the LHC with $\sqrt{s}=14$~TeV as a
function of the gluino mass. The CTEQ6L1 parton distribution
functions~\cite{Pumplin:2002vw} are employed, and the renormalization
and factorization scales are fixed at the average mass of the final
state particles, i.e. 
$\mu_R=\mu_F=(m_{\go}+m_{3/2})/2\sim m_{\go}/2$. As the cross
section scales as $m_{3/2}^{-2}$, we fix the gravitino mass here so
that the production cross section becomes comparable to the gluino pair
production process (shown by black lines). We also show 
contributions of each subprocess, the $gg$ and $q\bar q$ initial state,
with a dashed and dotted line, respectively. The $gg$ subprocess depends
only on the gluino mass once the gravitino mass is fixed, while the
$q\bar q$ initiated cross section also depends on the $t$- and
$u$-channel-exchanged squark masses. Here, the masses of the left-handed
and right-handed squarks are fixed at 3~TeV. It should be noted that
those contributions are not decoupled in the large squark mass, and the
heavier squark exchange increases the cross section since the
gravitino-quark-squark couplings are proportional to the squark mass
squared. Therefore, as one can see in figure~\ref{fig:xs_go_mass}, the
cross section of the $q\bar q$ channel can be larger than that of the
$gg$ channel even at the LHC.

\subsection{Gluino pair production}

In the scenario where the gravitino is the LSP and the gluino the NLSP,
gluino pair production gives rise to a di-jet plus missing energy
signature at the lowest order: 
\begin{align}
 pp\to\go\go\to gg\gro\gro.
\label{process_gogo}
\end{align} 
The LO cross section is shown in figure~\ref{fig:xs_go_mass} as a function
of the gluino mass.%
\footnote{In addition to the SUSY QCD interaction diagrams, there is the
$t$- and $u$-channel gravitino exchange contribution, which is, however,
negligible when 
$m_{3/2}>10^{-13}$~GeV~\cite{Dicus:1989gg,Drees:1990vj,Dicus:1996ua,Kim:1997iwa}.}
Unlike the $\go\gro$ associated production, the $\go\go$ production
needs the partonic energy to be at least twice the gluino mass, and
hence the cross section falls rapidly with the increase of the gluino
mass. For light gluinos the contribution from the $gg$ initial state is
dominant, while for heavy gluinos the production via the $q\bar q$
initial state becomes considerable. 

As one can see in figure~\ref{fig:xs_go_mass} with the fact of the
$m_{3/2}^{-2}$ scaling behavior of $\sigma(pp\to\go\gro)$, the different
gravitino and gluino masses alter the $n$-jet topology in the final
state. In other words, the kinematic distributions and the number of
jets in the final state might be able to give us some information on the
gluino mass and/or the gravitino mass. However, as mentioned in
section~\ref{sec:intro}, the detailed analysis of the multi-jet events
requires the ME+PS merging prescription. In the next section, therefore,
we will promote the previous LO
studies~\cite{Dicus:1989gg,Drees:1990vj,Dicus:1996ua,Kim:1997iwa,Klasen:2006kb} 
to a full-fledged simulation via a state-of-the-art event generator. 

Before turning to the ME+PS merging procedure, we briefly mention the
decay width of the NLSP gluino. The partial width of a gluino decay into
a gluon and a gravitino is given by 
\begin{align}
 \Gamma({\go\to g\gro})=\frac{m^5_{\go}}{48\pi\Mpl^2m^2_{3/2}},
\end{align}
where $\Mpl\equiv M_{\rm Pl}/\sqrt{8\pi}\sim2.4\times10^{18}$~GeV 
is the reduced Planck mass and the gravitino mass
in the phase space is neglected. For instance, for $m_{\go}=800$~GeV and
$m_{3/2}=3\times 10^{-13}$~GeV, the width is 4.1~GeV. In our simplified
SUSY mass spectrum the branching ratio is unity, $B(\go\to g\gro)=1$,
while one in the usual SPS7 and SPS8 GMSB benchmarks is discussed
in~\cite{Klasen:2006kb}. We remind the reader that the $\go\to g\gro$
decay is isotropic, and hence the gluon jet distribution is given by
purely kinematical effects of the decaying gluino.

\section{Merging matrix elements with parton showers}
\label{sec:merging}

In this section, we discuss the procedure used in this work to merge
matrix elements (ME) and parton showers (PS) for the
process~\eqref{signal_parton} as well as for the SM background, and show
the validation of our simulations. 

\begin{figure}
\center
 \includegraphics[width=.5\textwidth,clip]{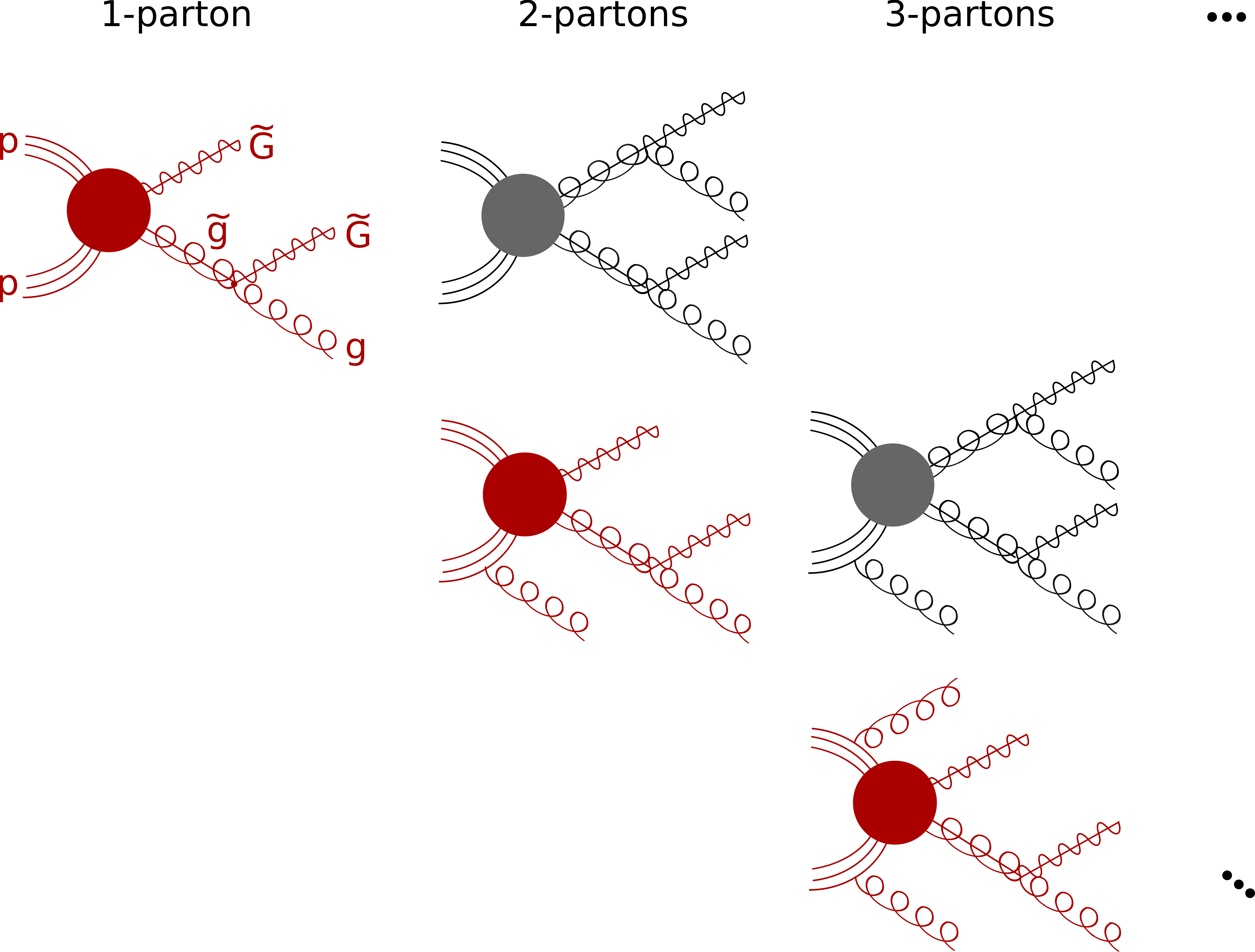} 
 \caption{Schematic diagrams for $pp\to\text{partons}+\gro\gro$. In the
 first row the leading gluino-gravitino (red) and gluino-pair (black)
 diagrams are sorted. The diagrams are ordered with the number of
 additional QCD partons in rows, while with the total parton
 multiplicity in columns.}
\label{fig:diagrams}
\end{figure}

At the LO, $\go\gro$ and $\go\go$ production are expected to lead
to missing energy in association with mono-jet and di-jet,
respectively. However, for production processes with large partonic
center-of-mass energy such as for heavy gluino production, initial and
final state QCD radiation becomes important, resulting in multi-jet
final states, and might modify or alter the LO predictions for the
relevant observables. In the present study, therefore, we consider
the processes beyond the LO ones, schematically presented 
in figure~\ref{fig:diagrams}. 

In general the signal may contain not only hard jets
from the decay of the gluinos as well as well-separated QCD radiation,
but also soft and/or collinear jets, which, if not properly treated,
lead to large logarithms. In event simulations, the hard partons are
described well by a fixed-order ME approach, while the soft and
collinear partons can be correctly described by a PS approach. 

To combine the two approaches avoiding double counting, one needs an
appropriate merging procedure. Several multi-jet merging algorithms have
been proposed (see also~\cite{Buckley:2011ms}): the CKKW-based 
method~\cite{Catani:2001cc,Lonnblad:2001iq}, the MLM
scheme~\cite{Mangano:2001xp,Alwall:2007fs}, the pseudo-shower
algorithm~\cite{Mrenna:2003if}, and the shower-$k_T$
scheme~\cite{Alwall:2008qv}. 

In our analysis we make use of the shower-$k_T$ scheme, which is based
on event rejection, as implemented in 
{\sc MadGraph}~\cite{Alwall:2007st,Alwall:2011uj} for fixed-order ME
generation and interfaced to {\sc Pythia6.4}~\cite{Sjostrand:2006za} for
PS and hadronization. In this scheme, ME multi-parton events are
generated with a minimum separation, $Q_{\rm cut}$ and 
$p_{T_{\rm min}}$, between final-state partons ($ij$) and between final-
and initial-state partons ($iB$) characterized by the $k_{T}$ jet
measure:   
\begin{align}
 d^2_{ij}=\min(p^2_{T_i},\,p^2_{T_j})\,\Delta R^2_{ij}>Q^2_{\rm cut},
 \quad d^2_{iB}=p^2_{T_i}>p^2_{T_{\rm min}},
\end{align}
with $\Delta R^2_{ij}=2[\cosh(\eta_i-\eta_j)-\cos(\phi_i-\phi_j)]$,
where $p_{T_i}$, $\eta_i$ and $\phi_i$ are the transverse momentum,
pseudorapidity and azimuth of particle $i$~\cite{Catani:1993hr}. The
renormalization scale for $\alpha_s$ for each QCD emission vertex is set
to the $k_T$ value, while the factorization scale for the parton
densities and the renormalization scale for the hard 2$\to$2 process is
given by the transverse mass of the particles produced in the central
process. The ME-level events are then passed to {\sc Pythia} and
showered using the $p_T$-ordered shower, and {\sc Pythia} reports the
scale $Q_{\rm hardest}^{\rm PS}$ of the hardest emission in the
shower. For lower parton-multiplicity samples an event is rejected if
$Q_{\rm hardest}^{\rm PS}>Q_{\rm cut}$, while for the highest
multiplicity sample an event is rejected if 
$Q_{\rm hardest}^{\rm PS}>Q_{\rm softest}^{\rm ME}$, the scale of the
softest ME parton in the event. See more details
in~\cite{Alwall:2008qv}.

\subsection{Physics parameters and observables}\label{parameters}

Throughout the present study, we consider a gluino with mass
$m_{\go}=800$~GeV, which lies above the exclusion limit for certain
simplified SUSY models or general gauge mediation
models~\cite{Aad:2011ib, Kats:2011qh}, and conduct analyses for the LHC
at $\sqrt{s}=14$~TeV. All the left- and right-handed squarks are fixed
at 3~TeV. The corresponding LO gluino-pair production cross section
$\sigma(\go\go)$ is about 1 pb at the 14-TeV LHC; see
figure~\ref{fig:xs_go_mass}. As discussed in detail in 
section~\ref{sec:go_gro_assoc}, the gluino-gravitino associated
production cross section $\sigma(\go\gro)$ strongly depends on the
gravitino mass. In the following we focus on three different
gravitino masses which exemplify the different final states. 
First, we fix the gravitino mass at $3\times 10^{-13}$~GeV so that
$\sigma(\go\gro)\sim\sigma(\go\go)$. We subsequently take a lighter
and a heavier gravitino as
\begin{subequations}
\begin{align}
 \text{A}\,(m^{}_{3/2}=1\times 10^{-13}\, {\rm GeV}): 
   &\qquad \sigma^{\rm A}(\go\gro) \sim 9\times\sigma(\go\go),\\
 \text{B}\,(m^{}_{3/2}=3\times 10^{-13}\, {\rm GeV}): 
   &\qquad \sigma^{\rm B}(\go\gro) \sim \sigma(\go\go),\label{XS_B}\\ 
 \text{C}\,(m^{}_{3/2}=9\times 10^{-13}\, {\rm GeV}): 
   &\qquad \sigma^{\rm C}(\go\gro) \sim \frac{1}{9}\times\sigma(\go\go).
\end{align}\label{ABC_masses}
\end{subequations}
Hence, $\go\gro$ associated production is dominant for case~A, while
$\go\go$ production is the main channel of the gravitino production for 
case~C. The two production processes are comparable in case~B. The LHC
may be able to explore the above mass range beyond the current
bound, eq.~\eqref{massbound}.  

We have fixed the above benchmarks based on the LO predictions for the
cross sections. It is well known, however,  that higher order QCD
corrections can enhance the expected rates.  
For instance, the next-to-leading order (NLO) cross
section for the gluino pair is 1.96 times larger than the LO cross
section for $m_{\go}=800$~GeV with $m_{\tilde q}=3$~TeV at the 14-TeV
LHC~\cite{Beenakker:1996ch}, while NLO corrections to $pp\to\go\gro$
have not yet appeared in the literature. We note that our analyses can
be easily redone with a different overall normalization and yet the main
features will not change. In any case our approach is complementary to a
fixed-order NLO calculation which reliably predicts cross sections and
observables involving at most one jet, while ME+PS merged computations
provide a reliable prediction for multi-jet based observables and more
exclusive quantities that can be directly used in experimental
simulations. 

Within the present study, the relevant observables are related either to
jets or missing energy. Here, we will focus on the following variables:
\begin{itemize}
 \item transverse momentum of the leading and second jets, 
       $p_T=|\vec p_T|$; 
 \item missing transverse energy, $\Etmiss$;
 \item sum of all the jet $p_T$'s, 
       $H_T\equiv\sum_j p_{T}^j$; 
 \item jet multiplicity.
\end{itemize}

\subsection{Technical setup for simulations}
\label{sec:technical_setup}

To simulate the signal process~\eqref{signal_parton}, we have
implemented the effective gravitino interaction Lagrangian~\eqref{L_int}
into {\sc FeynRules}~\cite{Christensen:2008py,Duhr:2011se}, which
provides the {\sc UFO} model file~\cite{Degrande:2011ua,deAquino:2011ub}
for ME generators.  
We use {\sc MadGraph5}~\cite{Alwall:2011uj} to generate the
ME multi-parton events both for the gravitino signal and the SM
background, and employ {\sc Pythia6.4}~\cite{Sjostrand:2006za} for PS
and hadronization. The shower-$k_T$ scheme is applied for the ME+PS
merging as described above. We have checked that all the ME-level
results as well as the merged results agreed with those by 
{\sc MadGraph/MadEvent v4}
with the gravitino code~\cite{Hagiwara:2010pi} and also the goldstino 
code~\cite{Mawatari:2011jy}.  

In the following analyses, we generate signal events with parton
multiplicity from one to three, $pp\to\gro\gro+1,2,3$ partons, and
merging separation parameters $Q_{\rm cut}=100$~GeV and 
$p_{T_{\rm min}}=50$~GeV. The choice of the merging parameters will be 
discussed in section~\ref{sec:validation}. Note that the employment of
the ME+PS merging scheme allows us to treat different contributing
processes (e.g. the $\go\gro$ and $\go\go$ production processes in our
case) within one event simulation and without double counting.

We also consider the irreducible $Z$+jets SM background,
$pp\to Z(\to\nu\bar{\nu})+1,2,3$ partons, with merging separation
parameters $Q_{\rm cut}=p_{T_{\rm min}}=30$~GeV. Simulation of the other
main background, e.g. $W$+jets and top pair, which requires more
dedicated analysis, is beyond the scope of the present study, and we 
refer to, e.g., \cite{Alwall:2008va} for details and to
\cite{Chatrchyan:2011zy,Aad:2011ib} for the experimental analysis.  

For the jet clustering, we employ {\sc FastJet}~\cite{Cacciari:2011ma}.
Jets are defined by the anti-$k_T$ algorithm~\cite{Cacciari:2008gp} with
a distance parameter of 0.5, and are required to satisfy $|\eta_j|<4.5$
and $p_{T_j}>50$~GeV. We order the clustered jets by their transverse
momentum.

\subsection{Validation}\label{sec:validation}

Although the above merging parameters have been chosen in accordance
with the guidelines in~\cite{Alwall:2008qv}, we have explicitly checked
the stability of the cross section with respect to the variation of the
arbitrary scale $Q_{\rm cut}$. 

\begin{figure}
\center
 \includegraphics[width=.48\textwidth,clip]{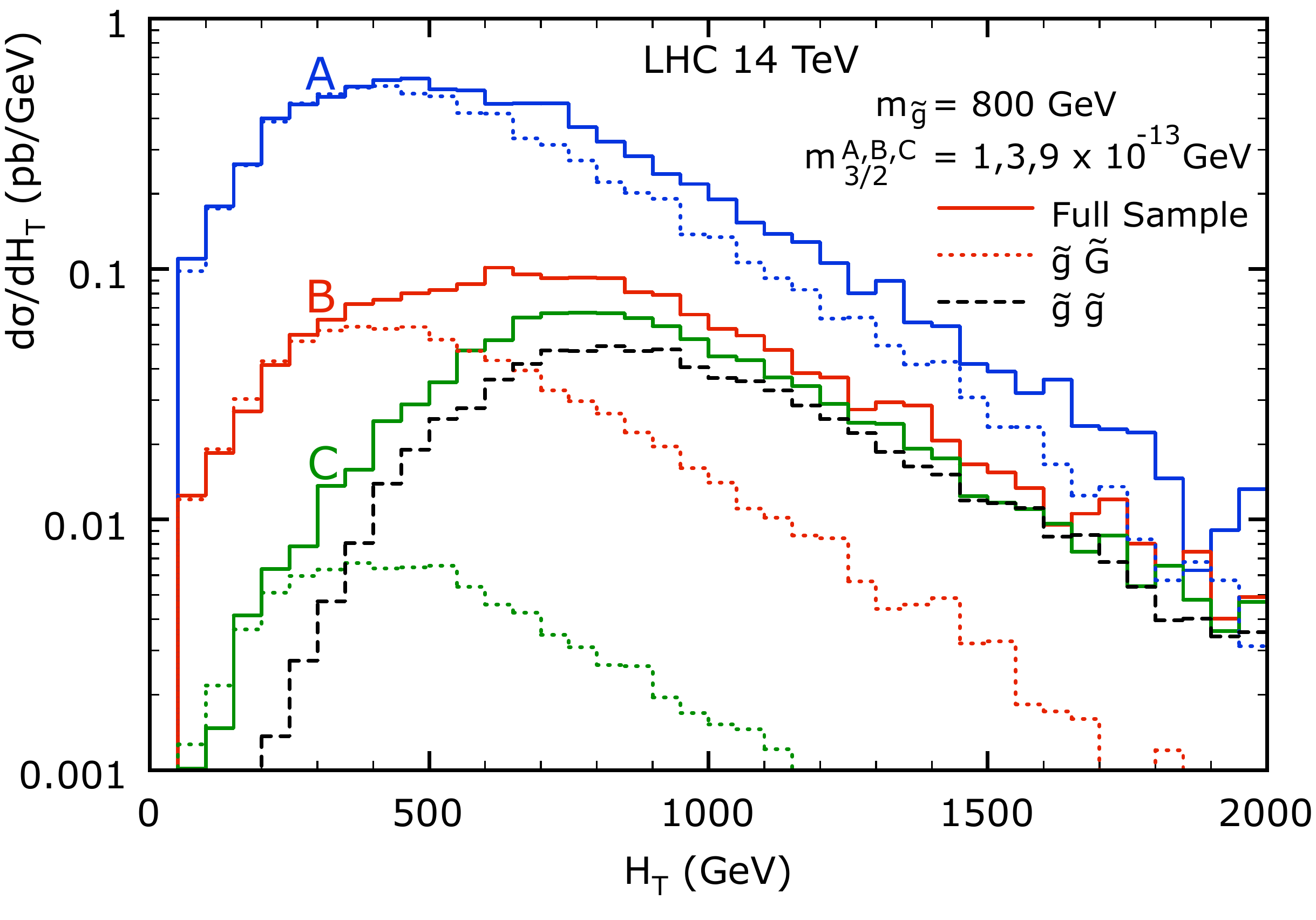}
 \quad
 \includegraphics[width=.48\textwidth,clip]{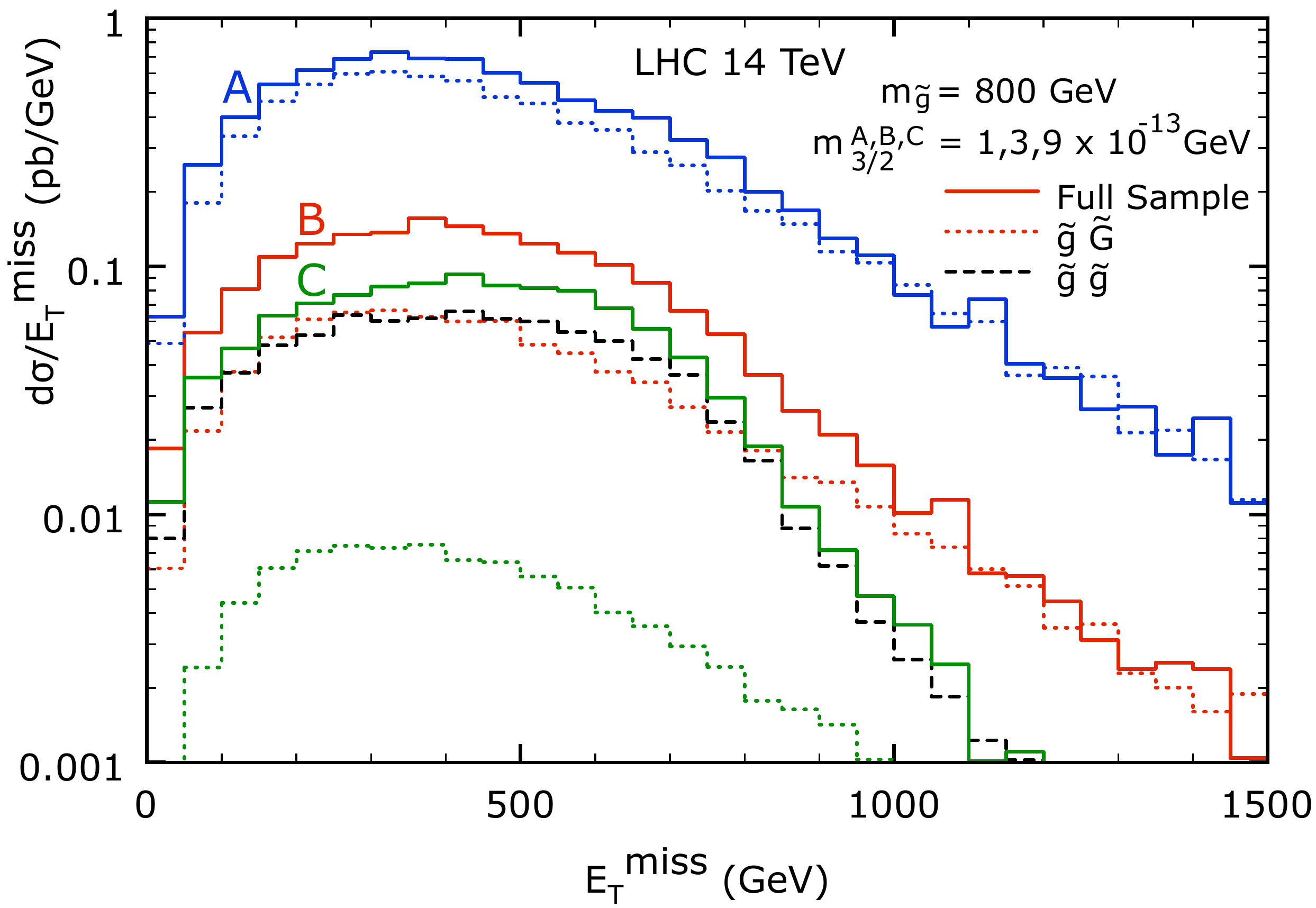}
 \caption{Shower $k_T$ merging results for inclusive signal samples of
 $pp\to{\rm jets}\,$+$\Etmiss$ at $\sqrt{s}=14$~TeV, where the gravitino
 mass is taken to be $m_{3/2}^{\rm A,B,C}= 1,3,9\times 10^{-13}$ GeV and
 the gluino mass is fixed at 800 GeV. The contributions of the
 gluino-gravitino associated production and the gluino-pair production
 are also separately shown by dotted and dashed lines, respectively.}
\label{fig:sum_ht}
\end{figure}

The smoothness of distributions across the transition between ME and PS
regimes was also examined for various $Q_{\rm cut}$ values and
kinematical distributions. Solid lines in figure~\ref{fig:sum_ht} show
the inclusive signal samples of $pp\to{\rm jets}$+$\Etmiss$ in the $H_T$ 
(left) and $\Etmiss$ (right) distributions. One can see the smooth
distributions for all the three benchmark points A, B, and C
in~\eqref{ABC_masses} for 
$m_{3/2}=1,3,\,{\rm and}\, 9\times 10^{-13}$~GeV, respectively.

In addition, as a nontrivial validation check, we have generated the
gravitino production subprocesses separately: 
$pp\to\go(\to g\gro)\gro+0,1$ partons and 
$pp\to\go(\to g\gro)\go(\to g\gro)+0,1$ partons, employing the same
merging procedure with the full signal sample, and verified that the
sum of those samples reproduces the full inclusive results. In
figure~\ref{fig:sum_ht}, we present contributions of each
subprocess, the $\go\gro$ production (dotted) and the $\go\go$
production (dashed). The sum of the two samples agrees with the full
samples (solid). We note that the cross section for the
$\go\gro$ production follow the $m_{3/2}^{-2}$ scaling, while the
$\go\go$ production is independent of the gravitino mass. 

For case B, as requested in~\eqref{XS_B}, the full signal cross section
consists of two equally relevant contributions coming from the $\go\gro$
and $\go\go$ production processes. In contrast, the signal of the
lighter gravitino (case A) is dominated by the $\go\gro$ associated
production process, and the signal for the heavier gravitino (case C)
consists mainly of the $\go\go$ production process.

The $H_T$ distributions for the $\go\gro$ production have a peak around
half of the gluino mass since there is a gluon coming from the gluino
decay, whose energy is $m_{\go}/2$ in the gluino rest frame. On the
other hand, the $\go\go$ production exhibits a peak around $m_{\go}$
due to the two gluino decays. 

The missing transverse energy $\Etmiss$ is defined as the absolute value
of the vectorial sum of the transverse momenta of the two gravitinos.  
The gluino-gravitino associated production leads to higher $\Etmiss$
events than the gluino-pair production, since a gravitino is directly
produced in association with a gluino and hence can have higher $p_T$
than the ones resulting from the gluino decays.

\begin{figure}
\center
 \includegraphics[width=.48\textwidth,clip]{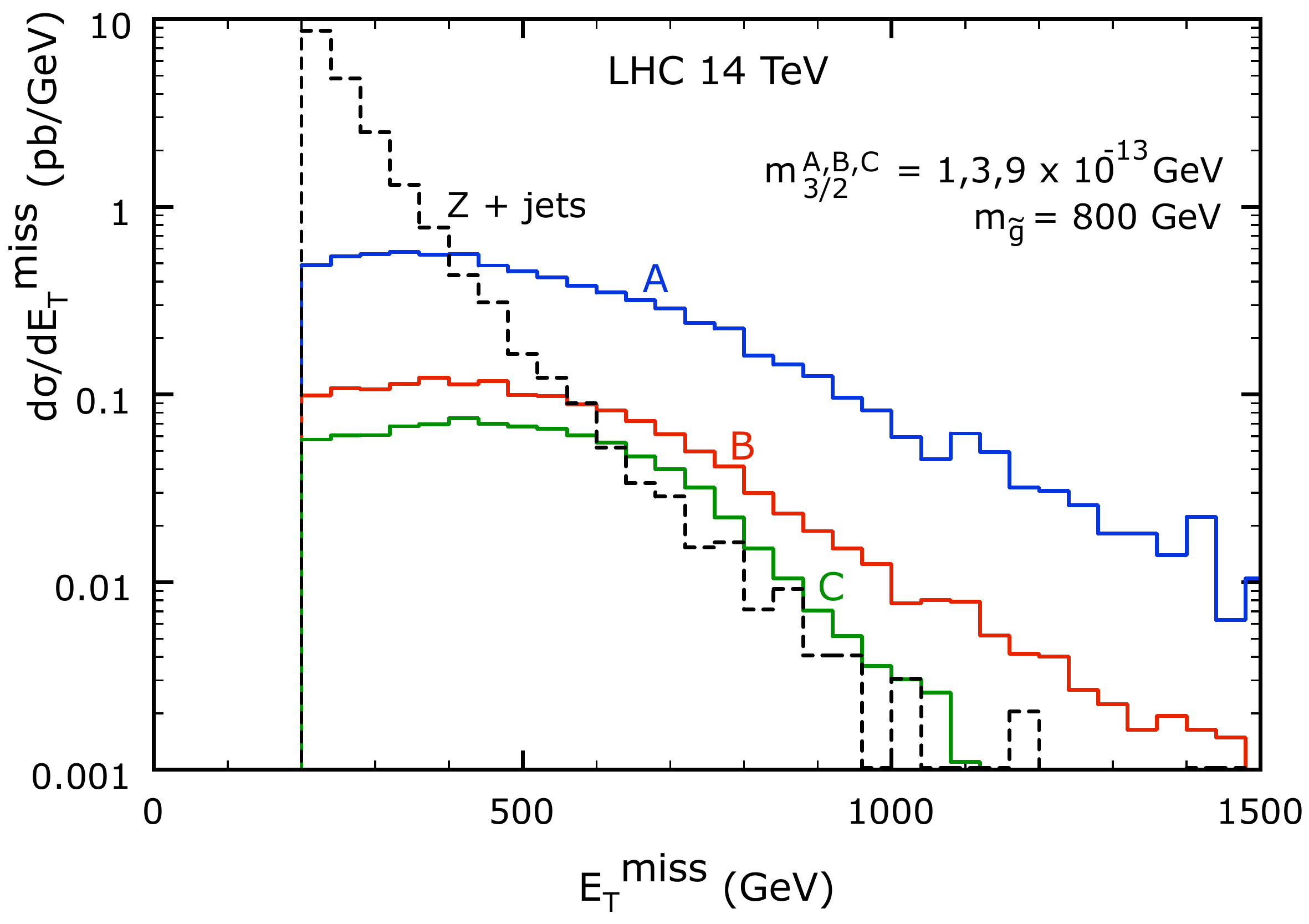}
 \caption{The same as the right plot in figure~\ref{fig:sum_ht} with the
 irreducible $Z(\to\nu\bar\nu)$+jets background (dashed), where the
 $\Etmiss>200$~GeV cut is imposed.}  
\label{fig:ResultsGld}
\end{figure}

Finally, we show the $\Etmiss$ distribution for the irreducible 
$Z(\to\nu\bar\nu)$+jets background in figure~\ref{fig:ResultsGld}, where
$\Etmiss=p_{T_Z}$. Since the background overwhelms the signal and
dominates in the low $\Etmiss$ region, we impose the minimal missing
transverse energy cut 
\begin{align}
 \Etmiss > 200\ {\rm GeV}
\label{etmisscut}
\end{align} 
in the following analyses.

\section{Jets plus missing energy}\label{sec:analysis}

We now investigate the kinematical distributions
further, focusing on the correlation between the $p_T$ of the leading
jet and the missing transverse energy, in order to differentiate our
three benchmark signals as well as to identify basic selection cuts
to curb the irreducible background. 

\begin{figure}[b]
\center
 \includegraphics[width=.325\textwidth,clip]{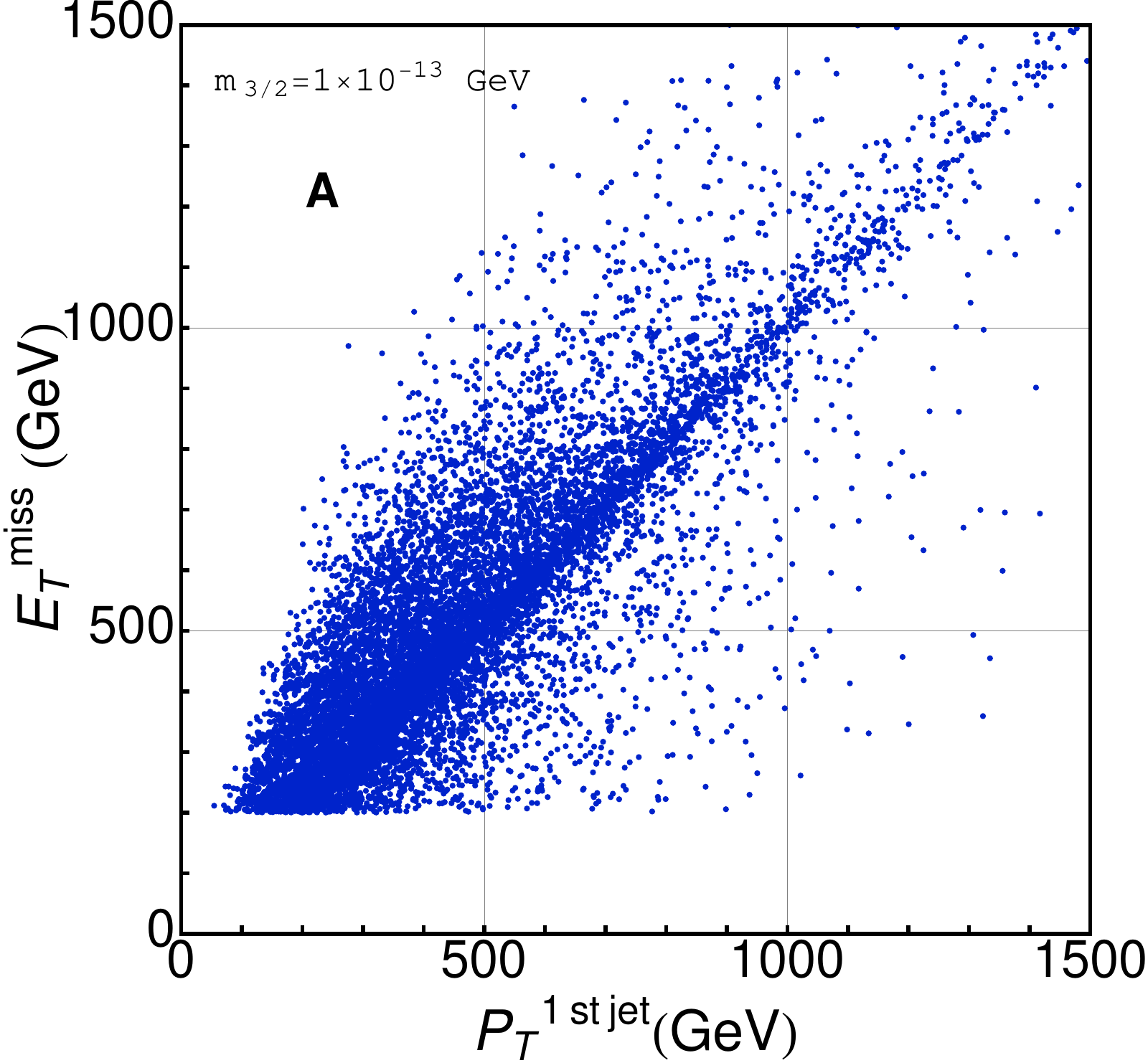}
 \includegraphics[width=.325\textwidth,clip]{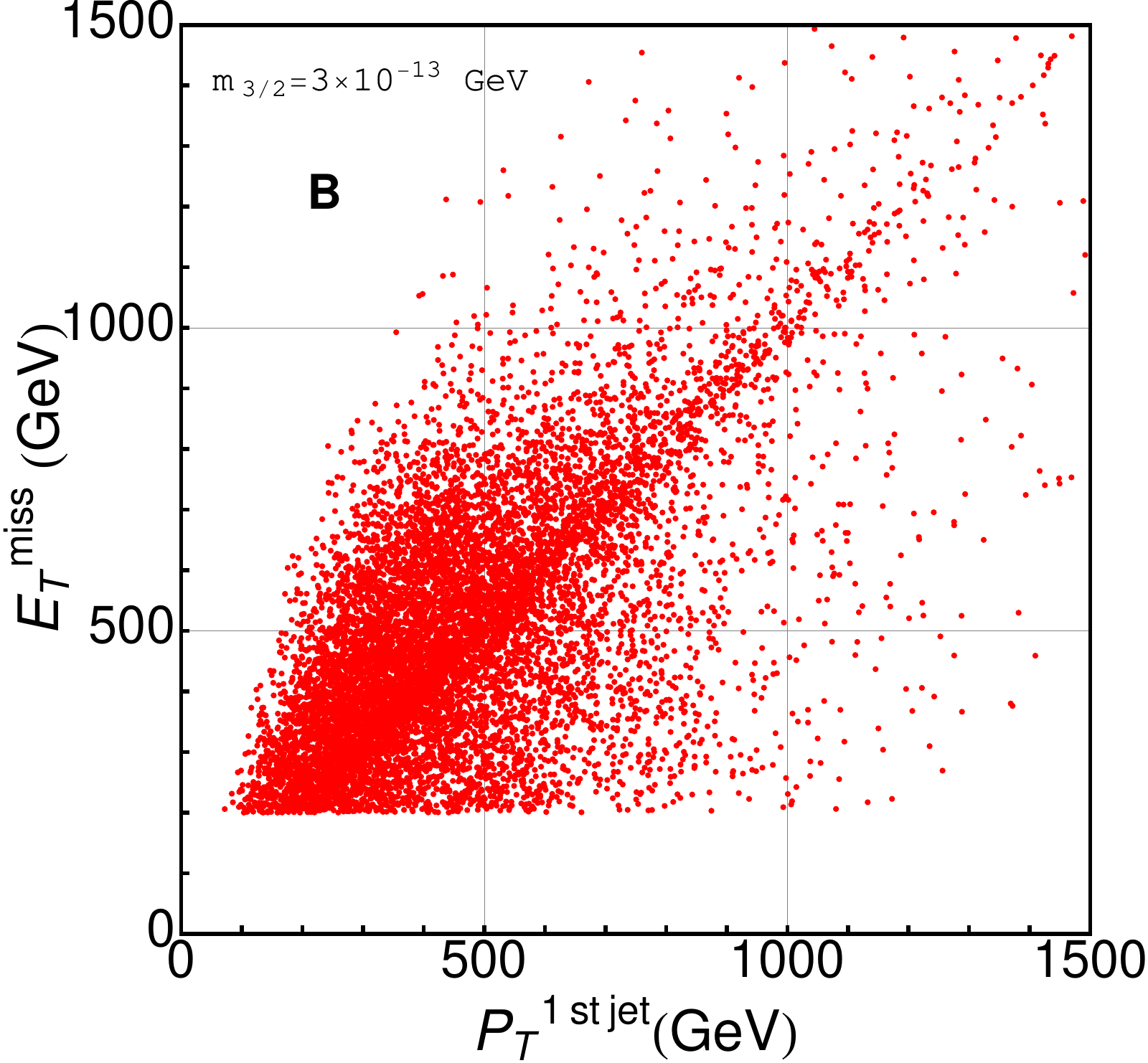}
 \includegraphics[width=.325\textwidth,clip]{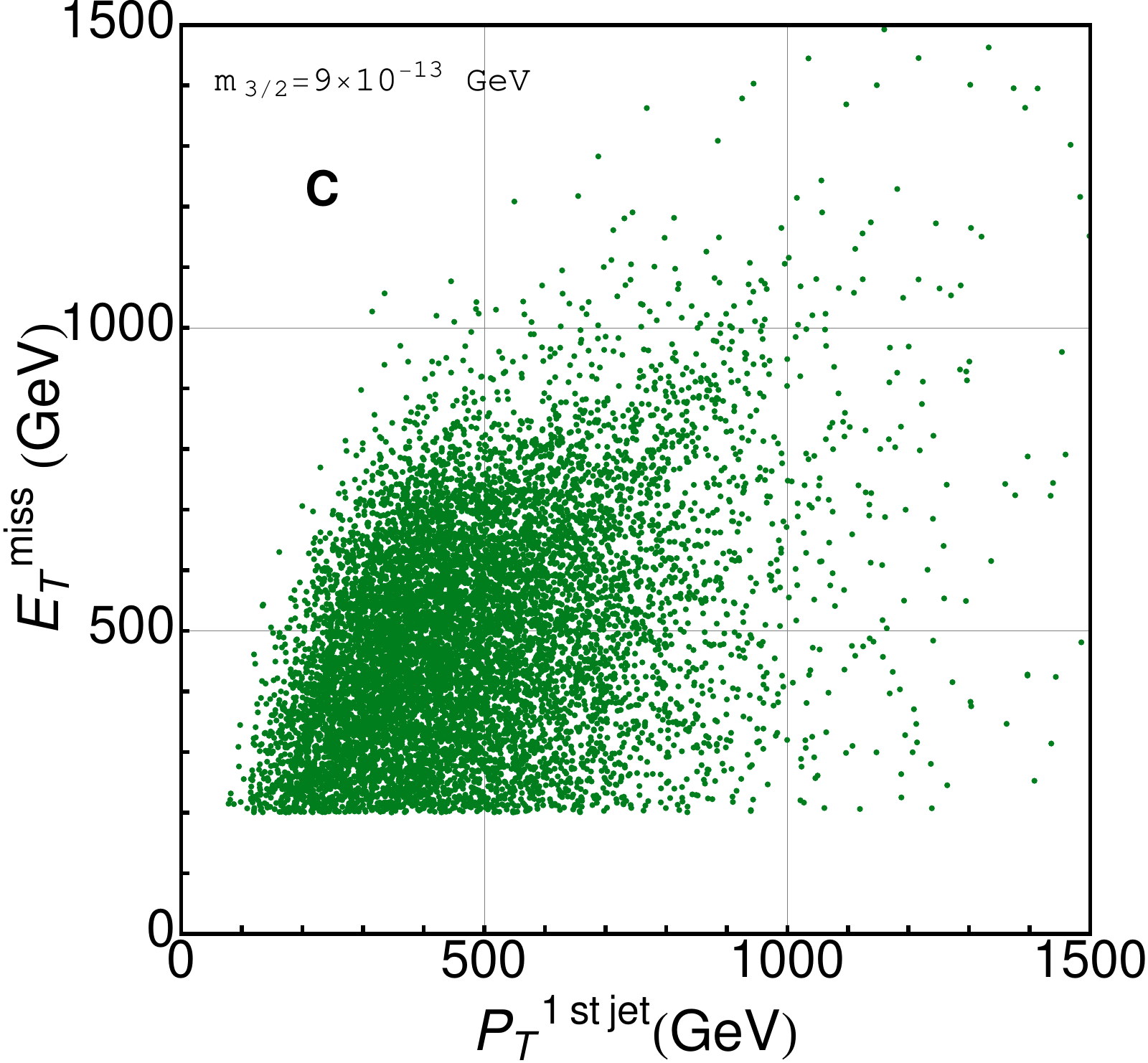}
 \caption{Scatter plots of the $pp\to{\rm jets}\ +\Etmiss$ signal at
 $\sqrt{s}=14$~TeV in the ($p_T^{\rm 1st\, jet},\,\Etmiss$) plane for
 $m_{3/2}^{\rm A,B,C}= 1,3,9\times 10^{-13}$ GeV from left to right,
 where the gluino mass is 800~GeV.} 
\label{fig:ScatterPlots}
\end{figure}
 
Figure~\ref{fig:ScatterPlots} presents scatter plots in the 
($p_T^{\rm 1st\, jet},\,\Etmiss$) plane for the three cases defined
in eqs.~\eqref{ABC_masses}, where the
minimal $\Etmiss>200$ GeV cut is applied. For case A,
where gluino-gravitino associated production is dominant, we find a
strong correlation between the two observables as 
$\Etmiss\sim p_T^{\rm 1st\,jet}$, especially for the high $p_T$ region,
and this can be explained as follows. One of two gravitinos in the final
state is produced in association with a gluino, and hence
$\vec p_{T_{\gro}}=-\vec p_{T_{\go}}$ at LO. The produced gluino
decays into a gluon and a (almost) massless gravitino, and those are
boosted along the gluino momentum direction and can share the momentum
like $\vec p_{T_g}\sim\vec p_{T_{\gro}}\sim\vec p_{T_{\go}}/2$. This
leads to a balance between the $p_T$ of the gluon jet and the missing
transverse energy, which is the vectorial sum of the two gravitino
momenta. QCD radiation will alter this naive expectation and most of
the events which scatter apart from the $\Etmiss=p_T^{\rm 1st\,jet}$
line come from samples with extra partons. 

For case C, in contrast, where gluino-pair production is the main
subprocess and both gluino decays are a source of the leading jet, there
is no such a strong correlation between $p_T^{\rm 1st\,jet}$ and
$\Etmiss$. In the high $p_T$ region, i.e. for the highly-boosted
gluino-pair production, a similar argument could be applied yet a
cancellation between the back-to-back gravitinos occurs, hence events
with large $\Etmiss$ are suppressed. This can be already observed in 
figure~\ref{fig:sum_ht}. Case B lies in between cases A and C, where
the both production subprocesses contribute.

\begin{figure}
\center
 \includegraphics[width=.325\textwidth,clip]{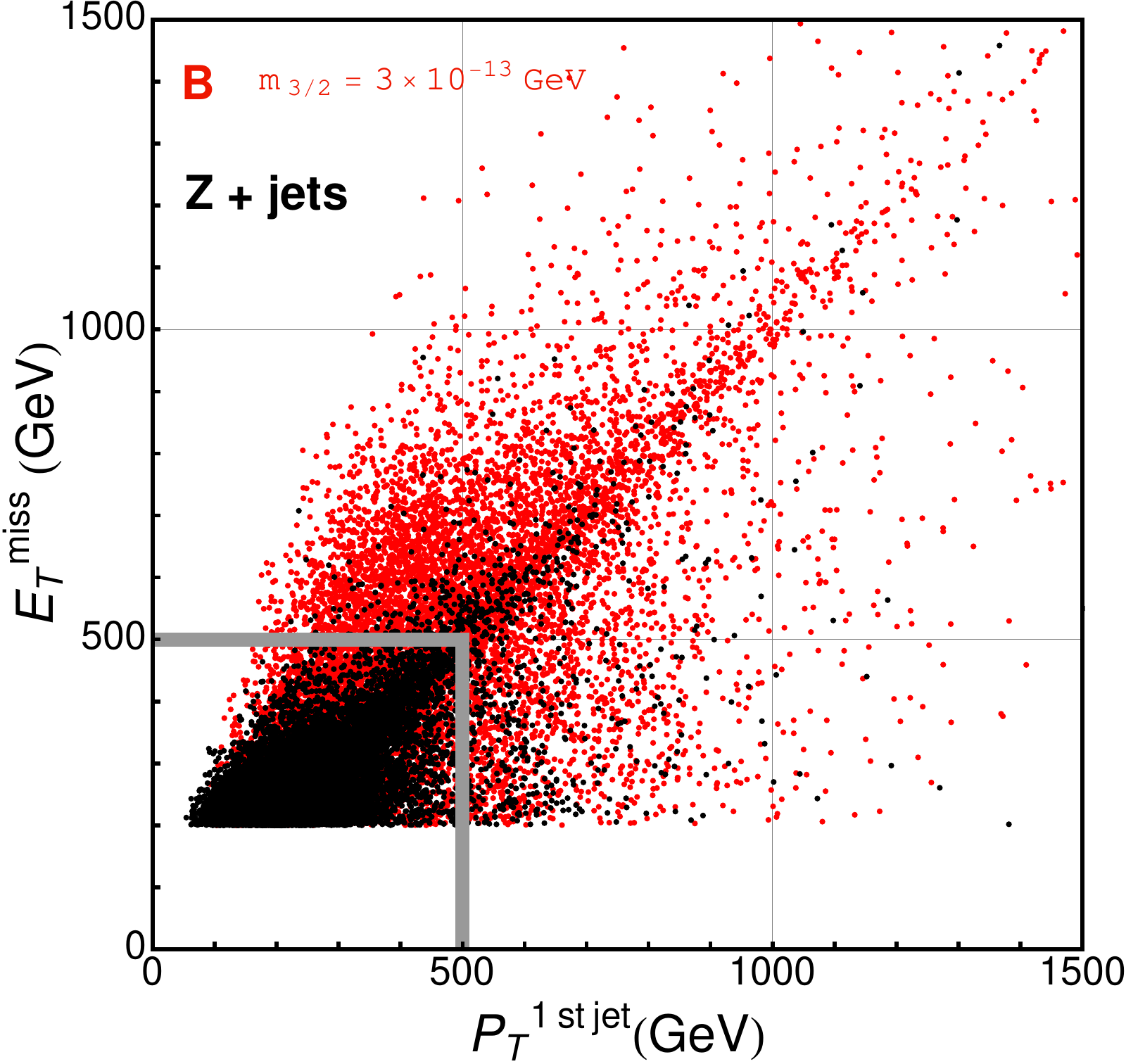}
 \caption{The same as the middle plot in figure~\ref{fig:ScatterPlots}
 with the $Z$+jets background (black dots).} 
\label{fig:ReduceBack}
\end{figure}

In figure~\ref{fig:ReduceBack} the SM $Z$+jets background is added on the
scatter plot for case B with black dots. Also here, we find (a weaker)
$\Etmiss\sim p_T^{\rm 1st\,jet}$ correlation resulting from the $Z+j$
sample. The background events are concentrated in the low $p_T$ and
$\Etmiss$ region, typically less than 500~GeV, while the gravitino
signal events are mainly scattered to the higher energy region up to
about 800~GeV, i.e. the gluino mass, as well as to the
$\Etmiss\sim p_T^{\rm 1st\,jet}$ region for cases~A and B. Therefore,
besides the minimal $\Etmiss$ cut in~\eqref{etmisscut}, we impose the
selection cuts 
\begin{align}
 p_T^{\rm 1st\,jet}>500\ {\rm GeV}\quad {\rm or}\quad
 \Etmiss>500\ {\rm GeV},
\label{eq:Cuts}
\end{align}
shown by thick grey lines in figure~\ref{fig:ReduceBack}.

\begin{table}
\center
 \begin{tabular}{|r||ccc|c|}
  \hline
  $\sigma$ (pb) & A & B & C & bkg  \\ \hline
  $\Etmiss>200$~GeV & 7.50 & 1.53 & 0.90 & 19.4 \\
  $+\ \ p_T^{\rm 1st\,jet}>500\ {\rm GeV}\ {\rm or}\
  \Etmiss>500\ {\rm GeV}$ & 3.81 & 0.85 & 0.55 & 0.81 \\
  \hline
 \end{tabular}
 \caption{Cross sections for the signals and the background at the
 14-TeV LHC, with the minimal $\Etmiss$ cut~\eqref{etmisscut} and with
 the additional selection cuts~\eqref{eq:Cuts}.} 
\label{table:xsec}
\end{table}

We present cross sections for the gravitino signals and
the $Z$+jets background in table~\ref{table:xsec}, where the minimal
$\Etmiss$ cut~\eqref{etmisscut} and the additional selection
cuts~\eqref{eq:Cuts} are taken into account. After the selection cuts,
the background is reduced quite effectively, while about half of the
signal events pass those cuts.  

\begin{figure}
\center
 \includegraphics[width=.48\textwidth,clip]{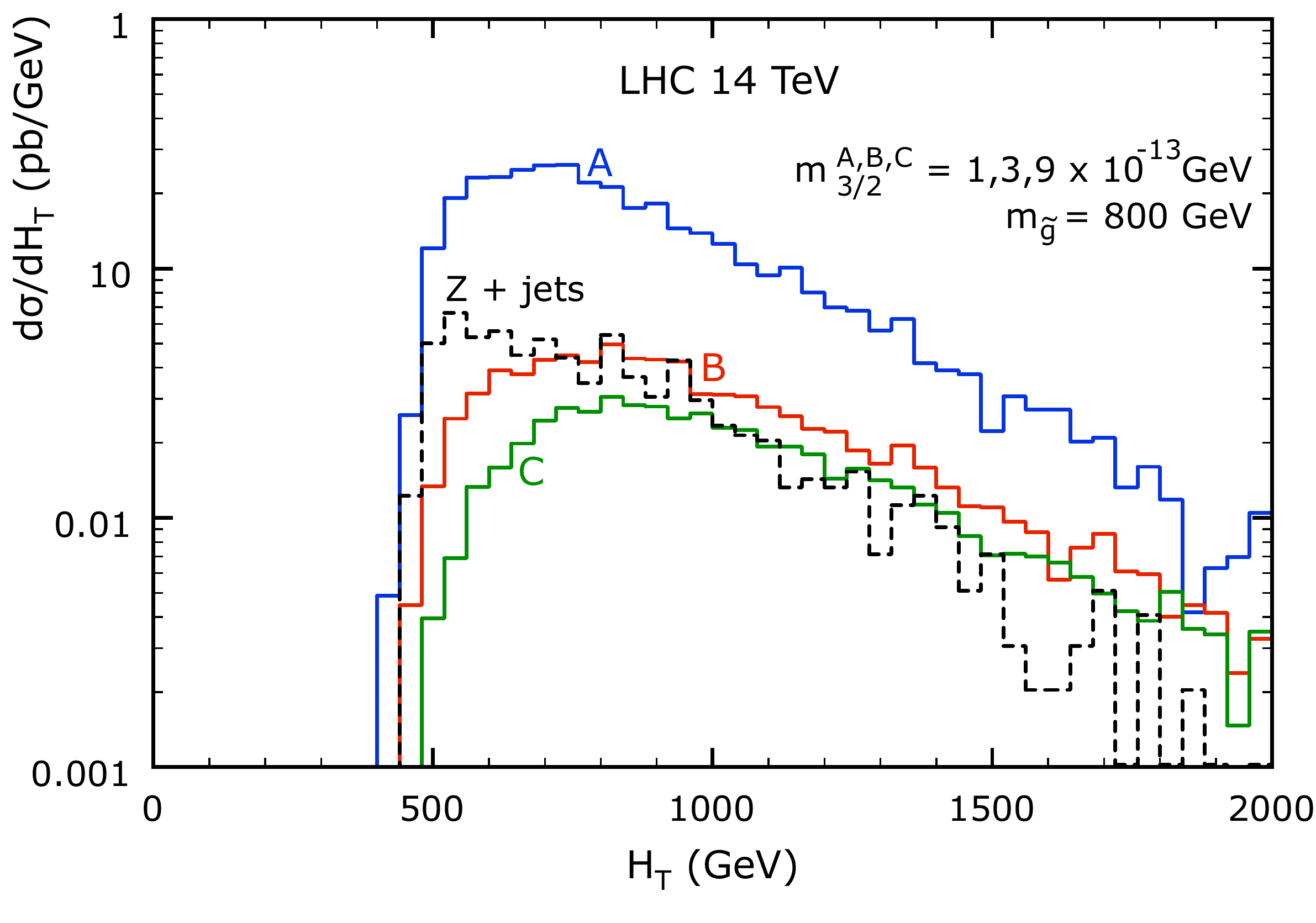}\quad
 \includegraphics[width=.48\textwidth,clip]{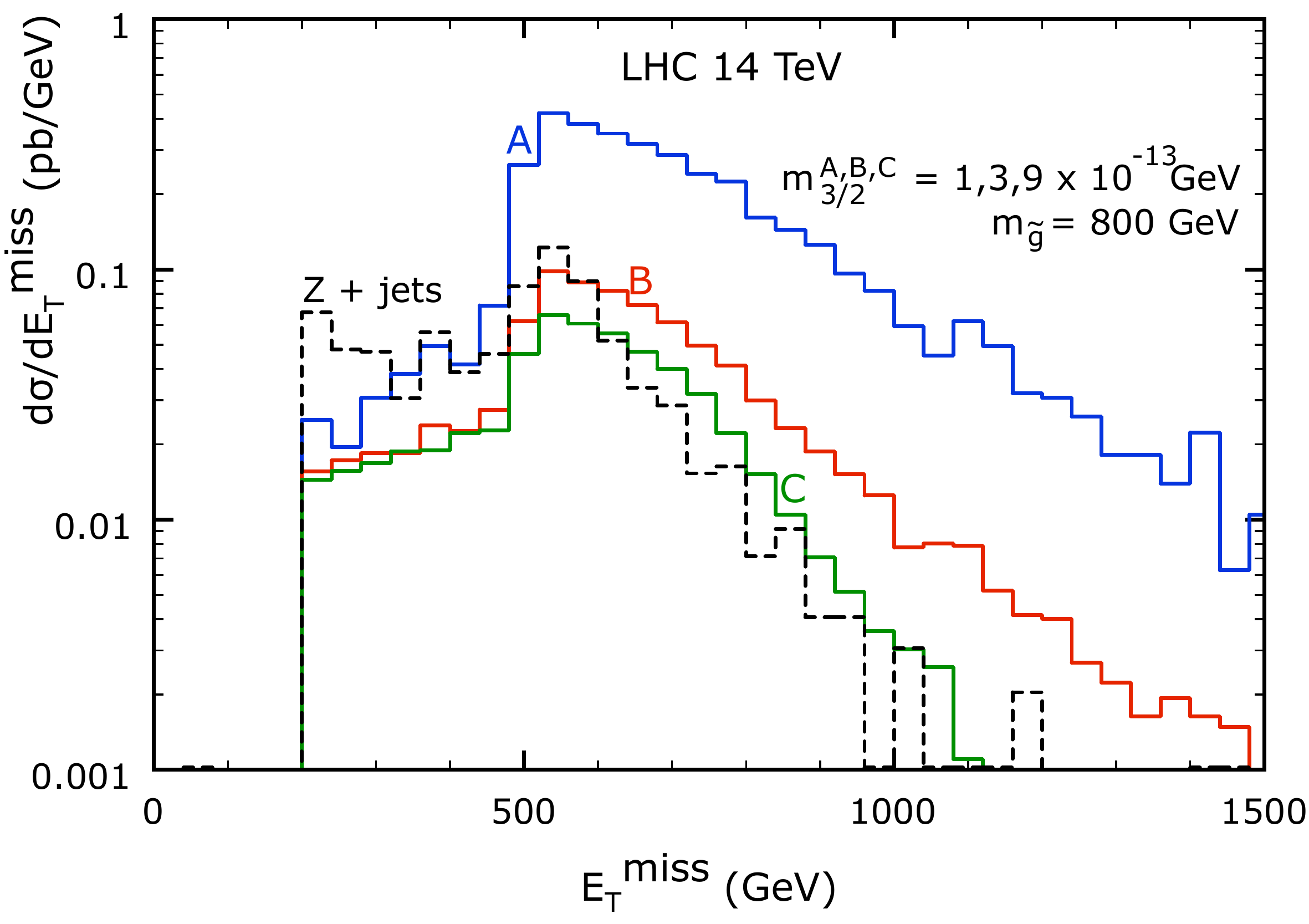} \\[3mm]

 \includegraphics[width=.48\textwidth,clip]{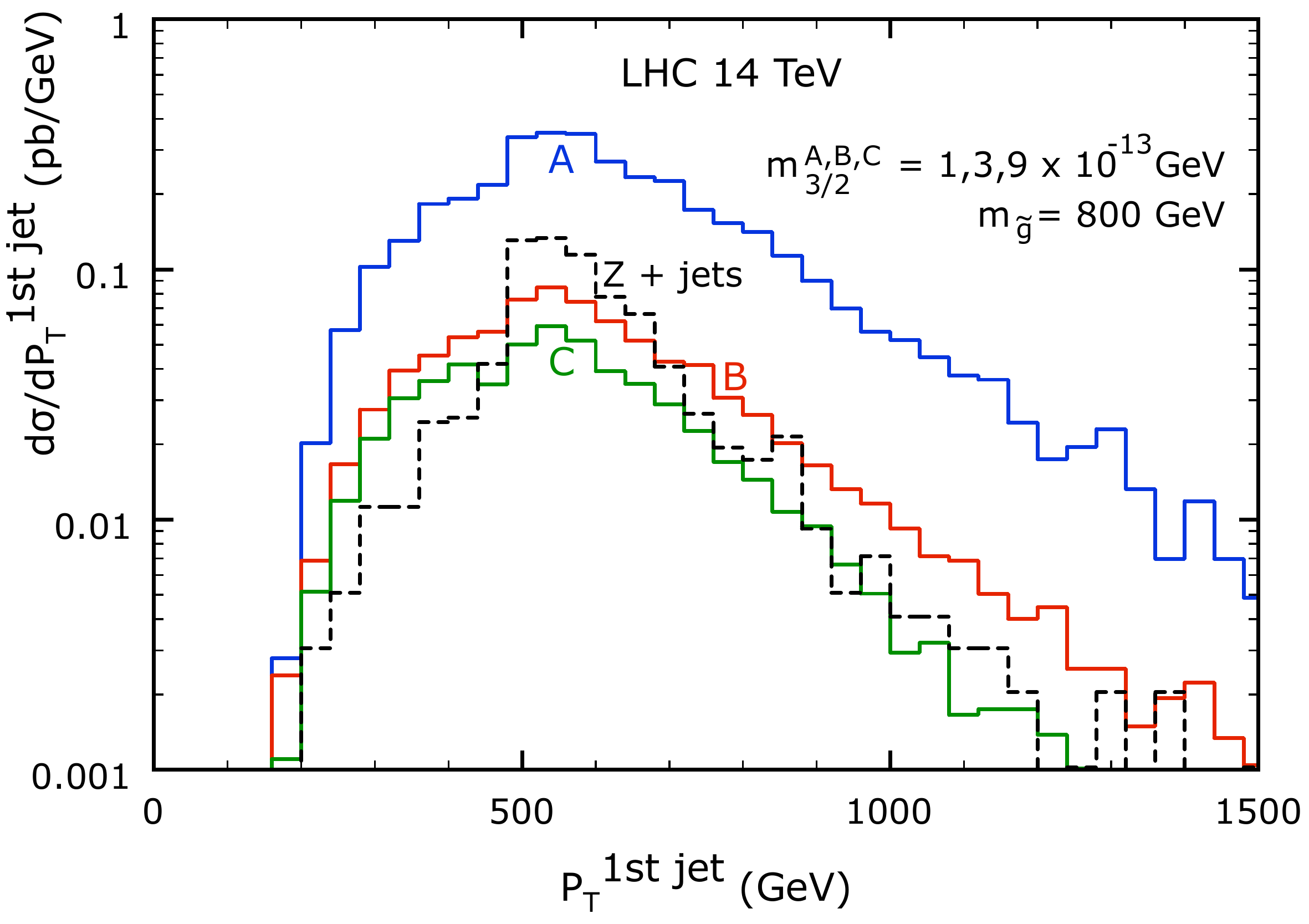}\quad
 \includegraphics[width=.48\textwidth,clip]{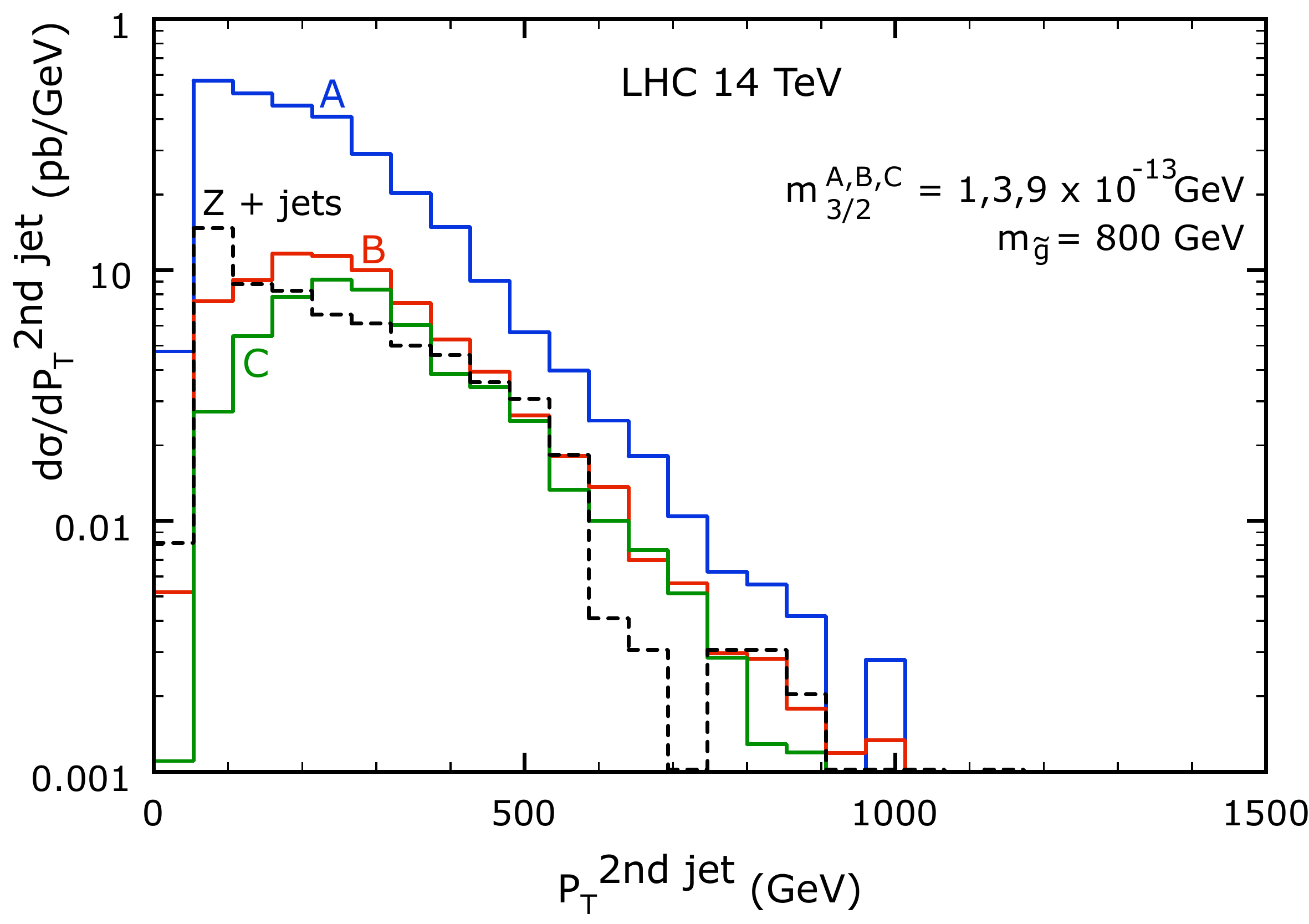}
 \caption{Distributions of the jets+$\Etmiss$ events at the 14-TeV LHC
 for the gravitino signal with $m_{3/2}^{\rm A,B,C}= 1,3,9\times
 10^{-13}$~GeV and $m_{\go}=800$~GeV as well as for the $Z$+jets
 background. Besides the minimal missing transverse energy cut
 $\Etmiss>200$~GeV, the selection cuts, $p_T^{\rm 1st\,jet}>500$~GeV or
 $\Etmiss>500$~GeV, are imposed.} 
\label{fig:dis}
\end{figure}

Distributions of the relevant observables given in
section~\ref{parameters} are collected in figure~\ref{fig:dis} for the
gravitino signals as well as the $Z$+jets background. Compared to
figure~\ref{fig:sum_ht}, events in the low $H_T$ and $\Etmiss$ regions
are removed by the kinematical cuts \eqref{etmisscut} and
\eqref{eq:Cuts}. In the missing energy distribution, as discussed above,
the lighter gravitino results in higher $\Etmiss$ events. 

The shapes of the $p_T$ of the leading jet are similar for the three
cases since the hard jets mainly come from the gluino decays, but the
$p_T$ distribution of the lighter gravitino case is slightly harder than
that of the heavier gravitino due to the higher boost effect from the
$\go\gro$ associated production. We also note that the signal events for
all the three cases dominate the background in the 
$p_T^{\rm 1st\,jet}<500$~GeV region. The distributions of the $p_T$ of
the second jet are more distinctive, especially in the low $p_T$
region. Two gluino decays in the gluino-pair production lead to two hard
gluon jets. On the other hand, the second jet resulting from the
$\go\gro$ production as well as the $Z$+jets background comes from QCD
radiation, and tends to be soft. 

\begin{figure}
\center
 \includegraphics[width=.48\textwidth,clip]{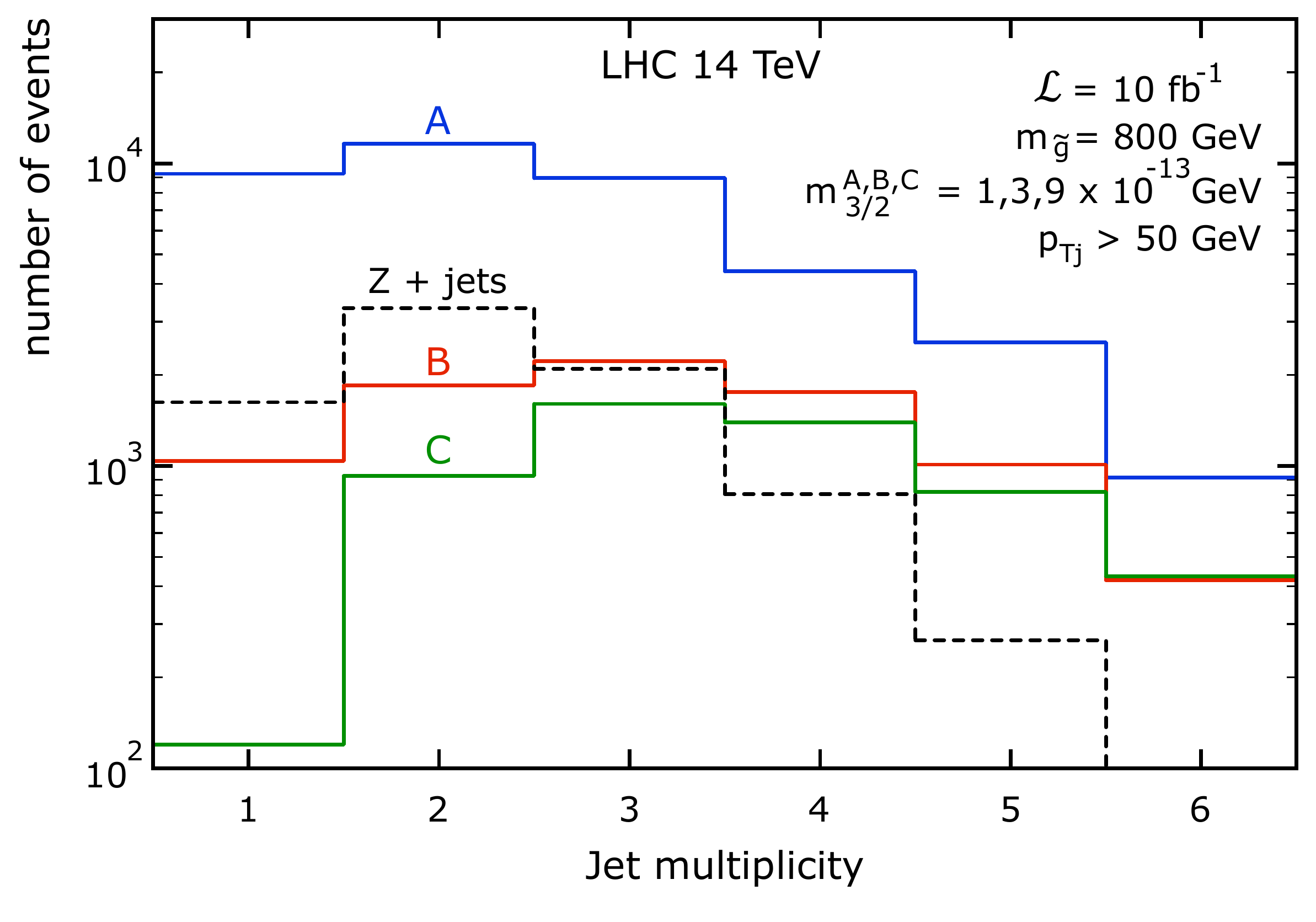}
 \quad
 \includegraphics[width=.48\textwidth,clip]{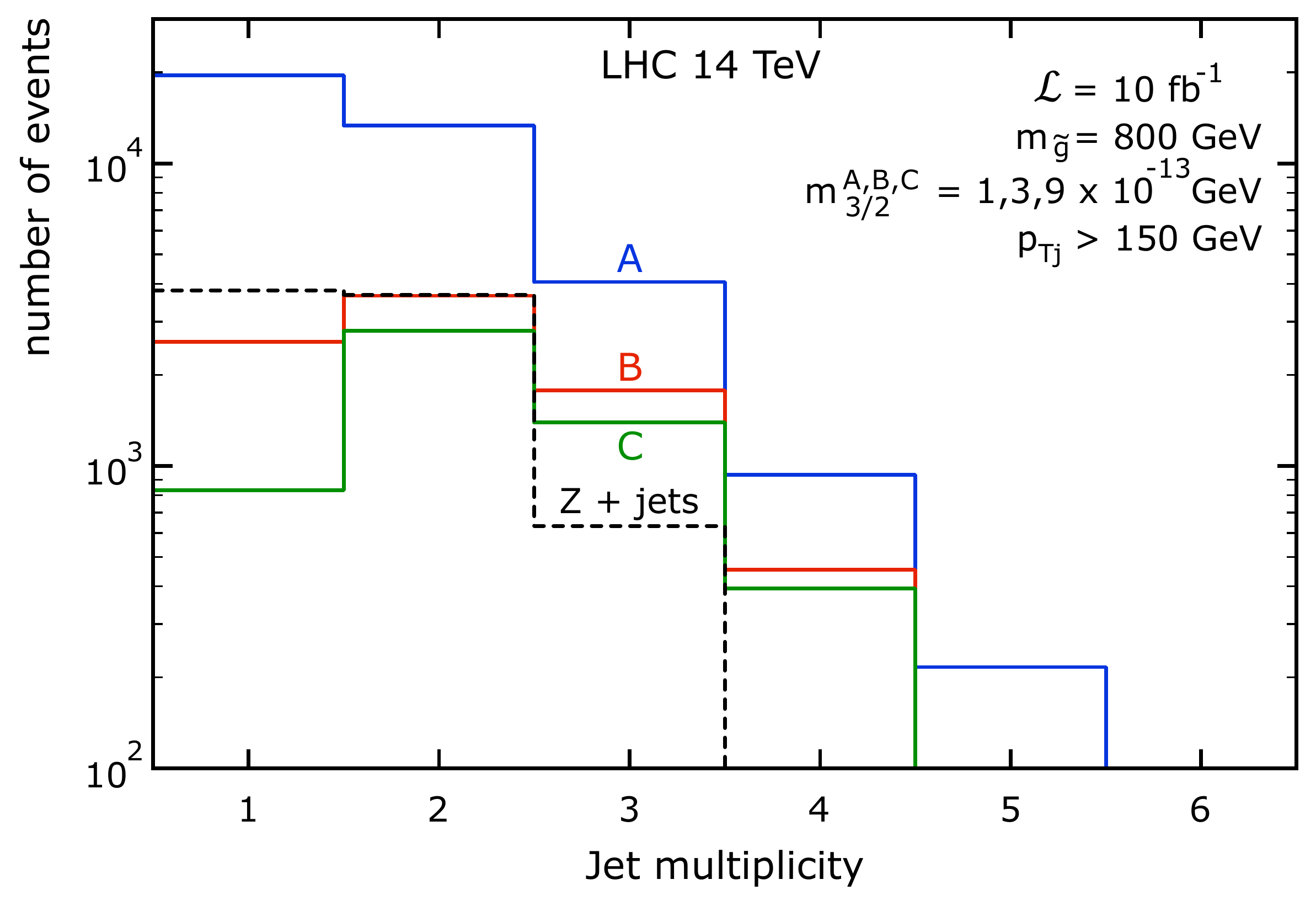}
 \caption{Jet multiplicities for an integrated luminosity of
 10~fb$^{-1}$, with $p_{T_j}>50$~GeV (left) and $p_{T_j}>150$~GeV
 (right). The detail is the same as figure~\ref{fig:dis}.}  
\label{fig:jetrate}
\end{figure}

Finally, we present jet multiplicities for an integrated luminosity of
$\mathcal{L}=10$~fb$^{-1}$ in figure~\ref{fig:jetrate}. The jet
multiplicity depends on the requirement of the minimal $p_T$ of jets,
and we take the different $p_{T_j}$ cuts of 50~GeV (left) and 150~GeV
(right). Case A as well as the SM background have a peak at a lower
multiplicity than cases B and C, as expected. When we count only jets
whose $p_T$ is larger than 150~GeV, i.e. only very hard jets,
distributions of the jet multiplicity recover the LO expectations: the
$\go\gro$ associated production tends to produce mono-jet events, while
the $\go\go$ production is likely to give di-jet events. 

As seen in figures~\ref{fig:dis} and \ref{fig:jetrate}, the
distributions are significantly different among the three benchmarks as
well as between the signal and the background. In other words, they are
sensitive to the gravitino mass when it is light enough so that the
$\go\gro$ associated production process can contribute to the signal. We
note that, although we fixed the gluino mass at 800~GeV in the present
study,  a different gluino mass also alters the distributions, which
could allow us to explore both the gravitino and gluino masses at the
LHC.

\section{Summary}\label{sec:summary}

We have studied a jets plus missing energy signature at the LHC in a
scenario where the gravitino is the LSP and the gluino is the NLSP which
promptly decays into a gluon and a gravitino. We considered a very light
gravitino of $m_{3/2}\sim{\cal O}(10^{-13}$~GeV), where two production
subprocesses can yield jets+$\Etmiss$: gluino-gravitino associated
production and gluino-pair production. By using the shower-$k_T$ ME+PS
merging scheme implemented in {\sc MadGraph}, we have simulated the
inclusive signal samples as well as the SM $Z$+jets irreducible
background. 

Special attention has been  devoted to the ME+PS merging procedure to
avoid double counting for such a signal which contains two different
types of subprocesses. In addition to checking the $Q_{\rm cut}$
independence of the cross sections and the smoothness of the
distributions, we have generated the merged $\go\gro$ and $\go\go$
signal samples separately and confirmed that the sum of them reproduced
the full inclusive results.

To show how distributions of the jets+$\Etmiss$ signature can provide
information on the gravitino and gluino masses, we have investigated
three benchmark scenarios which exemplify the different final
states. Due to the fact that the distributions are quite different
between the $\go\gro$ and $\go\go$ production processes and due to the
$m_{3/2}^{-2}$ scaling of the $\go\gro$ production cross section, the
kinematical distributions and the jet multiplicity exhibit distinctive
features among the three cases as well as between the signal and the
background. The LHC may be able to explore the parameter space around
our benchmark points and hence to provide information on the gluino mass
as well as the gravitino mass, yielding information on the SUSY breaking
scale.

\section*{Acknowledgments}

We would like to thank J.~Alwall, O.~Mattelaer and Y.~Takaesu for their
help with {\sc MadGraph}, and C.~Duhr and B.~Fuks for their help with
{\sc FeynRules}. 
This work is in part supported 
by the FWO - Vlaanderen, Project number G.0651.11, 
by the Federal Office for Scientific, Technical and Cultural Affairs
through the `Interuniversity Attraction Poles Programme' Belgian Science
Policy P6/11-P and VI/11, 
by the IISN MadGraph convention 4.4511.10,  
by the Concerted Research action
``Supersymmetric Models and their Signatures at the Large Hadron
Collider'' of the Vrije Universiteit Brussel (VUB), 
and by the VUB Research Council.

\appendix
\section{Effective gravitino interaction Lagrangian}
\label{sec:lagrangian}

We briefly present the relevant terms of the effective gravitino
Lagrangian in our study. In the high energy limit $\sqrt{s}\gg m_{3/2}$,
due to the goldstino equivalence theorem, the effective gravitino
interaction Lagrangian can be obtained by the replacement of the
spin-3/2 gravitino field ($\psi_{\mu}$) by the spin-1/2 goldstino field
($\psi$) as $\psi_{\mu}\sim\sqrt{2/3}\,\partial_{\mu}\psi/m_{3/2}$ in
the gravitino Lagrangian (see, e.g., eq.~(2) in~\cite{Hagiwara:2010pi}); 
see more details in~\cite{Mawatari:2011jy}. The effective interaction
Lagrangian among gravitino, quark and squark, $\psi$-$f$-$\phi$, and
among gravitino, gluino and gluon(s), $\psi$-$\lambda$-$A$(-$A$), in
non-derivative form is 
\begin{align}
  {\cal L}_{\rm int}
 =&\pm\frac{im_{\phi^i_{L/R}}^2}{\sqrt{3}\,\Mpl\,m_{3/2}}
   \big[\bar{\psi}P_{L/R}f^i(\phi^i_{L/R})^* 
  -\bar{f^i}P_{R/L}\psi\,{\phi}^i_{L/R}\big] \nn\\
  &-\frac{m_{\lambda}}{4\sqrt{6}\,\Mpl\,m_{3/2}} 
   \bar{\psi}[\gamma^{\mu},\gamma^{\nu}]
   \lambda^{a}F_{\mu\nu}^{a},
\label{L_int}
\end{align}
where $\phi_{L/R}$ denotes the left-/right-handed squark, 
$P_{L/R}=\frac{1}{2}(1\mp\gamma^5)$ is the chiral projection operator,
and 
$F^a_{\mu\nu}=\partial_{\mu}^{}A^a_{\nu}-\partial_{\nu}^{}A^a_{\mu}-g_sf^{abc}A^b_{\mu}A^c_{\nu}$
is the field-strength tensor of the $SU(3)_C$ gauge group
($a=1,\cdots,8$).

\end{document}